\documentclass[apj]{emulateapj}
\usepackage[colorlinks,linkcolor={blue},citecolor={blue},urlcolor={red}]{hyperref}
\usepackage{bm}
\usepackage{graphicx}
\usepackage{epsf}
\usepackage{graphics}
\usepackage{amsmath}
\usepackage{subfigure}

\usepackage{color, xcolor} 

\usepackage{soul} 
\soulregister{\cite}7 
\soulregister{\citep}{7} 
\soulregister{\citet}7 
\soulregister{\ref}7 
\def\beq{\begin{equation}}
\def\eeq{\end{equation}}
\def\bey{\begin{eqnarray}}
\def\eey{\end{eqnarray}}

\def\Mpc{\,{\rm Mpc}}
\def\mpc{\, h^{-1}{\rm {Mpc}}}

\def\mpci{\, {\rm {Mpc}}^{-1}}

\def\kms{\,{\rm {km\, s^{-1}}}}
\def\Msun{{\rm M_\odot}}

\def\Pre{\rm dyn~cm^{-2}}

\def\gs{\mathrel{\raise1.16pt\hbox{$>$}\kern-7.0pt
\lower3.06pt\hbox{{$\scriptstyle \sim$}}}}
\def\ls{\mathrel{\raise1.16pt\hbox{$<$}\kern-7.0pt
\lower3.06pt\hbox{{$\scriptstyle \sim$}}}}
\def\gtsima{\, {\buildrel > \over \sim} \,}
\def\ltsima{\, {\buildrel < \over \sim} \,}
\def\prosima{\, {\buildrel \propto \over \sim} \,}
\def\gsim{\lower.5ex\hbox{\gtsima}}
\def\lsim{\lower.5ex\hbox{\ltsima}}
\def\simgt{\lower.5ex\hbox{\gtsima}}
\def\simlt{\lower.5ex\hbox{\ltsima}}
\def\simpr{\lower.5ex\hbox{\prosima}}

\shorttitle{Shocks and Stripping}
\shortauthors{H. Li et al.}

\begin{document}
%\maketitle
% \linenumbers
% \switchlinenumbers %for line number
\title{Shock-induced stripping of satellite ISM/CGM in IllustrisTNG clusters at $z\sim0$}
\author{Hao Li\altaffilmark{1,2}, Huiyuan Wang\altaffilmark{1,2}, H. J. Mo\altaffilmark{3}, Yuan Wang\altaffilmark{1,2}, Xiong Luo\altaffilmark{1,2}, Renjie Li\altaffilmark{1,2}}
\altaffiltext{1}{Key Laboratory for Research in Galaxies and Cosmology, Department of Astronomy, University of Science and Technology of China, Hefei, Anhui 230026, China;lh123@mail.ustc.edu.cn,
whywang@ustc.edu.cn}
\altaffiltext{2}{School of Astronomy and Space Science, University of Science and Technology of China, Hefei 230026, China}
\altaffiltext{3}{Department of Astronomy, University of Massachusetts, Amherst MA 01003-9305, USA}
%\date{December 2021}
\begin{abstract}
Using the IllustrisTNG simulation, we study the interaction of large-scale shocks 
with the circumgalactic medium (CGM) and interstellar medium (ISM) of star-forming (SF) 
satellite galaxies in galaxy clusters. These shocks are usually produced by mergers 
and massive accretion. Our visual inspection shows that approximately half 
of SF satellites have encountered shocks in their host clusters at $z\leq0.11$. 
After a satellite crosses a shock front and enters the postshock region, 
the ram pressure on it is boosted significantly. Both the CGM and ISM can be severely 
impacted, either by striping or compression. The stripping of the ISM is particularly 
important for low-mass galaxies with $\log(M_*/\Msun)<10$ and 
can occur even in the outskirts of galaxy clusters. In comparison, satellites 
that do not interact with shocks lose their ISM only in the inner regions of 
clusters. About half of the ISM is stripped within about 0.6 Gyr after
it crosses the shock front. Our results show that shock-induced stripping 
plays an important role in quenching satellite galaxies in clusters.
\end{abstract}

%%%%%%%%%%%%%%%%%%%%%% SECTION 1%%%%%%%%%%%%%%%%%%%%%%%%

\section{Introduction}\label{sec_intro}

%First, many observational evidences show that environment play an important role in shaping galaxy properties, star formation, morphology and size. The dependence on halo mass, cluster-centric distance and galaxy morphology.
Galaxy properties are correlated with the environment in which they reside. 
It is well known that early-type galaxies are preferentially found 
in galaxy clusters \citep[e.g.][]{Dressler1980}, and late-type galaxies in clusters 
and fields are systematically different in their atomic and molecular gas contents
\citep{Solanes2001, Catinella2013, Boselli2014}. The fraction of 
quiescent galaxies is also found to increase with the local number density of galaxies, 
with halo mass, and with decreasing distance to halo center 
(halo-centric distance) \citep[e.g.][]{Baldry2006, Peng2010, 
Wetzel2013, Woo2013, WangH2018, Bluck2020}. Comparing satellite and central galaxies 
suggests that environmental effects on galaxies are complicated
\citep[e.g.,][]{vandenBosch2008, Wetzel2013, LiP2020} and depend 
strongly on intrinsic properties, such as galaxy stellar mass, 
morphology, and details of formation processes \citep[see e.g.][]{LiP2020}.
Various mechanisms have been proposed to explain the observed environmental 
dependence of galaxy properties, such as galaxy mergers and interactions
\citep{Toomre1972, Moore1996, Moore1998, Cox2006, Cheung2012}, 
tidal stripping \citep[TS, e.g.][]{Read2006}, strangulation 
\citep{Larson1980, vandenBosch2008, Peng2015}, and ram pressure stripping
\citep[RPS,][]{Gunn-Gott-72, Abadi-Moore-Bower-99, Boselli2014, Bosseli2022}.  

The ram pressure ($P_{\rm rp}$) on a galaxy depends both on the density of 
the ambient gas and on the velocity of the galaxy relative to the ambient 
gas, and so is expected to be strong in clusters of galaxies where 
the intra-cluster medium (ICM) is dense and typical velocity of 
member galaxies is high. The effectiveness of the RPS on the gas associated 
with a galaxy depends also on how strongly the gas is bound by the galaxy.   
Thus, the RPS has often been invoked for galaxy evolution in clusters, 
particularly for low-mass satellite galaxies, where the inter-stellar medium 
(ISM) and the circum-galactic medium (CGM) may be stripped relatively easily. 
Direct evidence for the RPS comes from the observed asymmetric gas distribution, 
particularly as long cometary tails that do not contain significant 
amounts of stars \citep{Bosseli2022}. Such stripped tails have been detected 
in both the inner regions and outskirts of nearby clusters and rich groups
\citep[e.g.,][]{Gavazzi1995, Kenney2004, Sun2007ApJ, Smith2010, Ebeling2014, Jachym2014, Poggianti2017ApJ, Chen2020, WangJ2021}. 

Based on galaxy properties and ICM density profiles obtained from
observational data and simulations, analytic models have been developed to 
estimate the efficiency of the RPS and its dependence on both galaxy and cluster properties
\citep[e.g.][]{Gunn-Gott-72, Hester2006, Koppen2018, Bosseli2022, VegaMartinez2022}. 
For a late-type galaxy of stellar mass $M_*=10^9\sim10^{10}\Msun$, it is 
found that the ram pressure can effectively strip the ISM at its stellar effective radius 
when $P_{\rm rm}\sim 10^{-12} \rm dyn~cm^{-2}$, and that 
the ISM can be fully stripped only when the satellite falls into the inner region 
of a cluster \citep[see e.g.][]{Bosseli2022}. For massive late-type galaxies 
($M_*>10^{10}\Msun$), a much higher ram pressure, $P_{\rm rm}\sim 10^{-11} \rm dyn~cm^{-2}$ 
is required for a significant stripping of the ISM.
RPS models based on these results have been implemented in some semi-analytic 
galaxy formation models to explain observed properties of satellite galaxies
\citep[e.g.][]{Guo2011, Henriques2015}.

The details of the RPS can be better understood using hydrodynamic simulations. 
To explore the dependence on various model parameters, controlled simulations are 
often employed, in which galaxy properties, galaxy orbits and inclinations, 
and ICM properties are set up by hand \citep[][]{Farouki1980, Abadi-Moore-Bower-99, 
Quilis2000, Marcolini2003, Roediger2005, Roediger2007, Haan2014, 
Roediger2014, Steinhauser2016, Steyrleithner2020}. The simulation results are 
broadly consistent with analytic models, although there are significant differences 
in details. These simulations demonstrate that ram pressure can produce a long 
cometary gas tail without significantly affecting stellar components. 
Sometimes the ram pressure can even compress the ISM at the leading edge 
and in the central region of stripped galaxies, thereby enhancing the star 
formation activity. The typical timescale for gas stripping in dwarf 
galaxies is about 100-500 Myr, while massive spirals can only be severely 
stripped in some extreme cases.

There are also attempts to investigate the effects of RPS using cosmological 
hydrodynamic simulations, where the properties of satellites 
mimic those of real galaxies and the evolution is followed in 
a realistic ICM \citep[e.g.][]{Tonnesen2008, Bahe2015, Lotz2019, Arthur2019, Yun2019}. 
It is found that the efficiency of the RPS can be suppressed or enhanced 
depending on different circumstances. For instance, some satellites 
may fall into clusters along with the IGM in filaments and have 
their gas protected by the filament gas until they reach the inner 
regions of clusters \citep{Kotecha22}. In such cases, the ram pressure 
must first get rid of the IGM and CGM associated with satellites, 
before it can strip the ISM. It was also found that feedback from 
active galactic nuclei (AGNs) in central galaxies can significantly change 
the ICM, consequently affecting the efficiency of RPS \citep{Martin-Navarro2021}. 
Moreover, the RPS is also sensitive to the presence of substructures 
in a cluster, where the ICM density is expected to be higher \citep{Tonnesen2008}.
The RPS efficiency also depends on satellites themselves. For example, ram pressure 
usually compresses the ISM and enhances star formation for gas-rich 
satellites \citep{Troncoso-Iribarren2020},  and internal processes, such as 
stellar feedback, may make the ISM fluffy and more susceptible to stripping
\citep{Bahe2015}. 

It is well known that many galaxy clusters are undergoing or have 
undergone merging processes of massive accretion. The collision of the ICM 
with the infalling gas and mergers can induce strong shocks 
\citep[e.g.][]{Kang2007ApJ, Bykov2008SSRv, Zinger2018MNRAS, LiR2022}, 
which are the hallmark of ICM environments. These shocks can generate 
jumps in gas pressure, density, temperature, and velocity, thus 
leading to significant enhancements in the ram pressure. 
\cite{Roediger2014} used controlled simulations to study the impact of ICM shocks 
on a massive spiral galaxy, and found that shocks can effectively strip its 
ISM. However the investigation focused on whether or not the interaction with 
shocks can produce star-forming tails, rather than on gas stripping. 
Some indirect observational studies provide support to the role 
played by shocks in stripping the ISM. For example, NGC 4522 has an asymmetric gas 
distribution, which is hard to explain using the RPS of a static ICM 
and hints the importance of shocks \citep{Kenney2004}. Signatures of RPS 
are expected to be more common in merging clusters where shocks are 
generated \citep[e.g.][]{McPartland2016, Stroe2017, Ebeling2019, Ruggiero2019}. Consistent with this,  
stripping signatures are observed in several galaxies around merger shocks in 
A1367 \citep[see discussion in][]{Bosseli2022} and Abell 2744\citep[e.g.][]{Owers2012}.

In this paper, we use hydrodynamic simulations in IllustrisTNG to study whether or 
not shocks in galaxy clusters have significant impacts on satellite galaxies. 
In particular, we focus on the stripping of CGM and ISM of satellites.
Our goal is to understand the role of RPS in quenching star formation of cluster galaxies. 
In Section \ref{sec_md}, we introduce the simulations, cluster and galaxy samples, and 
the method to identify events of shock-satellite interaction. 
Section \ref{sec_re} shows the effects of large-scale shocks on the CGM and ISM of 
satellites, their dependence on stellar mass and halo-centric distance, and results 
on stripping timescale. We compare these satellites with those that do not interact 
with large-scale shocks. Finally, we summarize and discuss our results in 
Section \ref{sec_sum}.

\section{Method}\label{sec_md}

\begin{figure*}[htb]
    \centering
    \includegraphics[width=.9\textwidth]{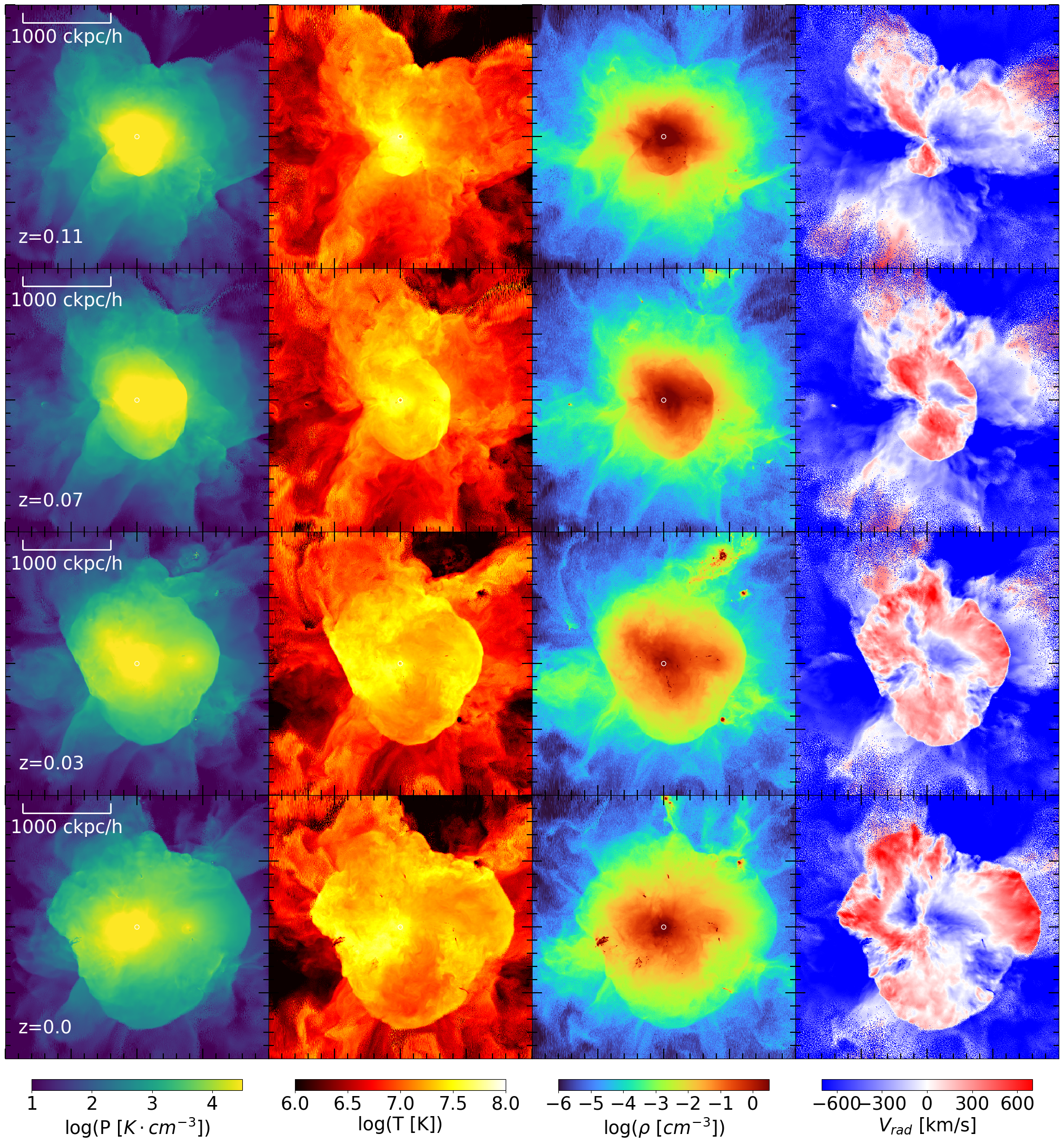}
    \caption{The propagation of a strong shock in gas pressure, temperature, density and radial velocity maps in a TNG cluster with $\log(M_{200}/\Msun)=14.4$. The white circle indicates the center of the cluster. A bow-like shock propagates from the cluster center to the right side, and an accretion shock appears to the left of the cluster.}
    \label{fig_shock} 
\end{figure*} 

\subsection{The IllustrisTNG Simulation}\label{sec_sim}

We investigate the interaction between shocks and satellite galaxies using data 
from the IllustrisTNG simulations \citep[e.g.][]{Naiman2018, Pillepich2018b, Nelson2018, Springel2018}, 
run with the AREPO code \citep{Springel2010}. 
The gravitational forces are evaluated using a particle-mesh and oct-tree 
algorithm, and the solutions to hydrodynamic equations are obtained by 
using a finite-volume Godunov method on a moving unstructured mesh defined by 
Voronoi tessellations of discrete points. The code thus captures the advantages of both 
Lagrangian and Eulerian methods. A series of tests presented in \cite{Springel2010} 
and \cite{Schaal2015} show that the moving-mesh technique can resolve shocks 
that are difficult to resolve in SPH-based techniques (see also Figure \ref{fig_shock}). 
TNG simulations include key physical processes for the formation and
evolution of galaxies, such as gas cooling and heating, star formation in a multiphase 
ISM, stellar population evolution, metal enrichment of gas, and feedback from both 
supernovae and supermassive black holes. The details of these physical processes are 
described in the two method papers \citep{Weinberger2017, Pillepich2018}. 
The simulations can reproduce the observed bimodal color distribution of galaxies, 
the fraction of quiescent galaxies and their dependence on galaxy stellar mass and 
environment \citep[e.g.][]{Nelson2018, Donnari2021}. This indicates that key 
processes related to quenching of star formation, such as AGN and stellar feedback, 
ram pressure stripping, tidal stripping, are modeled realistically in these simulations.

The cosmological parameters for the TNG simulations are consistent with recent Planck 
measurements \citep{Planck2016}: $\Omega_{m} = \Omega_{dm} + \Omega_{b} = 0.3089$, $\Omega_{b} = 0.0486$, 
$\Omega_{\Lambda} = 0.6911$, $\sigma_8=0.8159$, $H_{0} = 100h \kms\mpci$ with $h = 0.6774$ and $n_{s} = 0.9667$.
A series of simulations were carried out by TNG. To obtain a large sample of 
satellites with sufficiently high mass resolution, we decide to use the highest-resolution 
run of the $\sim100\Mpc$ simulations, TNG100-1 \footnote{https://www.tng-project.org/}, 
one of the three flagship simulations of the IllustrisTNG project. It follows the 
evolution of $1820^3$ dark matter (DM) particles and approximately $1820^3$ gas cells 
from $z=127$ to $z=0$. The masses of DM particles and gas cells are 
$\sim7.5\times10^{6}\Msun$ and $\sim1.4\times10^{6}\Msun$, respectively.
One hundred snapshots between $z=20$ and $z=0$ are dumped. There are 10 snapshots
between $z=0.11$ and $z=0$, the redshift range we are interested in. 
The time interval between two adjacent snapshots ranges from 0.130 to 0.204 Gyr, 
much smaller than the dynamic timescales of halos.

Halos and subhalos used here are obtained using the Friends-of-Friends (FoF) 
\citep{Davis85} and SUBFIND algorithms \citep{Springel2001,Dolag2009}. Galaxies are 
defined as the baryonic component in subhalos. The most massive subhalo in an FOF halo is 
classified as the central subhalo, and its galaxy is regarded as the central galaxy
of the halo. The baryonic components in other subhalos are referred to as satellite galaxies. 
The subhalo and galaxy merger trees are constructed over the 100 snapshots by the SUBLINK 
algorithm \citep{Rodriguez-Gomez2015}. In the following, we use galaxy merger trees to 
follow the motion of satellites in galaxy clusters and trace their evolution.

The halo mass, $M_{200}$, used in this paper is defined as the mass contained
in a sphere centered at the halo center, within which the mean mass density equals
200 times the critical density of the Universe. The radius of the sphere is referred to 
as the virial radius, $R_{200}$. The galaxy stellar mass, $M_*$, is the sum of all stellar 
particles within $2r_{\rm e}$, where $r_{\rm e}$ is the stellar half mass radius. 
We are particularly interested in the stripping of the CGM and ISM. We define the ISM 
of a galaxy as the cold gas cells ($\log(T/{\rm K})<5$) within 2$r_{\rm e}$. For CGM 
we focus on the gas medium within a spherical shell between $2r_{\rm e}$ and $4r_{\rm e}$. 
The details of the CGM and ISM quantities are given in the relevant sections.

\subsection{Galaxy clusters, large-scale shocks and satellite samples}\label{sec_sample}

In the TNG run used here, there are 14 galaxy clusters with $\log(M_{200}/\Msun)>14.0$ 
at $z=0$. We investigate satellite galaxies in these clusters, their interaction with ICM 
shocks, and the stripping and compression of the CGM and ISM.
Clusters are usually growing rapidly, and may be undergoing recent massive mergers. 
Our inspection shows that the fourteen clusters experienced 8 mergers with merger mass ratios 
above $0.1$ at $z\leq 0.2$. As shown in previous studies \citep{Ryu2003, Zinger2018MNRAS}, 
cluster mergers and accretions of filament gas can induce strong shocks, with scales 
comparable to the clusters themselves. 

Figure \ref{fig_shock} shows
the propagation of a large-scale shock in a cluster with $\log(M_{200}/\Msun)=14.4$ 
and $R_{200}=1.33\Mpc$ at $z\sim0$. 
From the evolution of the radial velocity map, we can see that, at
$z=0.11$, the gas stream with a large infall (blue) velocity plunged  
into the very center of the cluster. \cite{Zinger2018MNRAS} 
demonstrated that such a deeply penetrating stream in galaxy clusters 
can generate multiple strong shocks. Indeed, we see that bow shocks are 
generated in the cluster center and propagating outwards. It is clear 
that the post-shock region has higher pressure, density and temperature than 
the pre-shock region, and that the two regions have very different velocity 
structures. The shock surface, where jumps in gas pressure, temperature, density 
and velocity are located, is sharp and can be identified easily.

\begin{figure*}[htb]
    \centering
    \includegraphics[width=.8\textwidth]{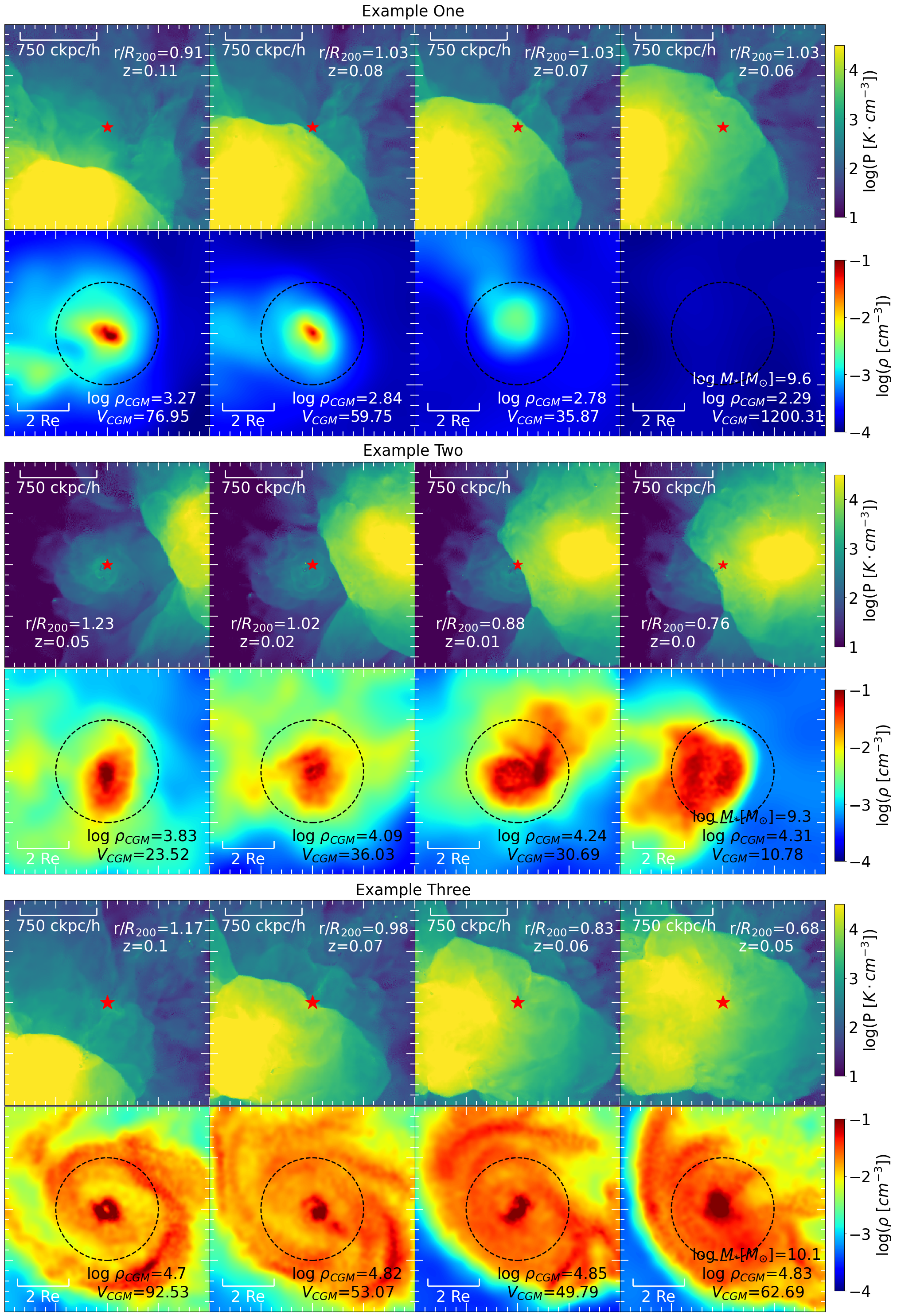}
    \caption{Three ShSAT examples (examples one, two and three) and the evolution of the medium in and around these satellites. For each example, the first row shows the gas pressure map at a large scale (cluster size) and the second row shows the gas density at a scale comparable to the galaxy size.  
    The panels in the second column show the maps at $t_{\rm sh}$ when the satellites encounter the shock surfaces. In each panel, we show $\rho_{\rm CGM}$, $v_{\rm CGM}$, the halo-centric distance ($r/R_{200}$) and redshift ($z$). The red stars in the first rows mark the position of the galaxies and the black circles in the second rows have radii of $2r_{\rm e}$.}
    \label{fig_examples} 
\end{figure*}

\begin{table}
\centering
\caption{Number of satellite galaxies in different subsamples}\label{table1}
%\begin{threeparttable}
%\resizebox{\textwidth}{!}
{
\begin{tabular}{c c c c}
\hline\hline
Mass range$^a$ & ShSAT & NsSAT & All \\
\hline
[9.0, 10.0) & 94 & 91 & 185\\\relax
[10.0, $\infty$) & 28 & 31 & 59 \\\relax
[9.0, $\infty$) & 122 & 122 & 244 \\
\hline
\hline
\end{tabular}}
%\begin{tablenotes}
%\footnotesize
\\
$^a$ The stellar mass ($\log (M_*/\Msun)$) range\\
%\end{tablenotes}
%\end{threeparttable}
\end{table}

These large-scale shocks cover a large fraction of the sky viewed from the galaxy clusters, 
and so they must interact with satellite galaxies as they propagate. 
It is then interesting to identify satellites that have interacted with these shocks.
To avoid potential evolution effects of galaxy clusters and satellites, we only consider 
interactions that occurred recently, i.e. at $z\leq 0.11$. 
We only consider galaxies of $M_{*}\geq10^{9}\Msun$, which have 
at least 1000 star particles and can be well resolved by the simulation. 
Since we are interested in satellite galaxies, we only select galaxies that have been 
identified as satellites at least once in its history and have ever reached 
a distance less than $1.5R_{200}$ from the centers of their host clusters at 
$z\leq0.11$. To avoid satellites that have been severely influenced by environmental processes 
before interacting with shocks, we require that, at $z=0.11$, the satellite resides in the 
outskirt of its host cluster, i.e. at a distance $r>0.8R_{200}$ from the central galaxy.  
We also discard satellites that are quenched ($\log\rm sSFR\leq 10^{-11}\ yr^{-1}$) 
or have a low gas fraction (with the ISM to stellar mass ratio $<5\%$) at $z=0.11$. 
Because we want to trace the evolution of the CGM and ISM of satellite galaxies in 
detail, the merger trees of satellites are required to be complete at $z\leq0.11$.
We discard seven satellites whose merger trees miss at least one snapshot. 

For each satellite, we generate gas pressure, density, temperature 
and velocity maps in $2\times2\mpc$ slices with a thickness of 
$0.15\mpc$ centered on the satellite. 
To avoid uncertainties due to projection effects, we inspect 
each map in three different ($X-Y$, $Y-Z$ and $X-Z$) planes. 
We trace the maps of the four properties from $z=0.11$ to 0, 
and visually identify whether or not the satellite has ever crossed a 
shock surface at $z\leq 0.11$. In most cases, the shock crossing event 
is easy to identify visually. We show three examples of the crossing events 
in Figure \ref{fig_examples}. As long as a satellite has crossed a shock surface, 
it is referred to as a ShSAT. Otherwise the satellite is referred to as a NsSAT. 
NsSATs are used to compare with ShSATs. Note that these large-scale shocks 
are not produced by satellites themselves. So in general a ShSAT moves from 
a pre-shock region to a post-shock region, as shown in Figure \ref{fig_examples}.

Due to the time resolution between snapshots, we cannot determine accurately 
when and where a shock-crossing event occurs. We thus choose the 
snapshot just before a satellite enters the post-shock region 
as the shock-crossing time, and we denote it by $t_{\rm sh}$. The distance 
of the satellite to the central galaxy at $t_{\rm sh}$ is denoted by $r_{\rm sh}$. 
To study the effect of shocks on satellites, we also consider two additional times
corresponding to the two snapshots before and after the snapshot of $t_{\rm sh}$, 
respectively, each of which is separated from the $t_{\rm sh}$ by two snapshots.   
These two times are denoted as $t_{\rm b}$ and $t_{\rm a}$, respectively. 
A few satellites cross the shock surface in the last two snapshots of the simulation and thus 
have no $t_{\rm a}$. They are excluded from our analysis. 
{And as shown in Section \ref{sec_re}}, if a satellite meets a shock at a large radius, 
the shock usually only has a weak impact on the satellite, we thus also discard 
shock-crossing events with $r_{\rm sh}>1.5R_{200}$. 
Finally, we obtain a total of 122 ShSATs and 122 NsSATs, as listed in 
Table\,\ref{table1}. The fact that the numbers of ShSATs and NsSATs are 
comparable suggests that shock-crossing is very common for satellites in 
clusters of galaxies. 

There are three reasons for us to rely on visual inspections to identify shock-crossing events of satellites. 
First, as shown {in Figure \ref{fig_shock}}, shocks in simulated clusters are usually 
very sharp and are easily recognizable by eye. Second, the shock crossing 
can sometimes distort the shock surface, making it difficult to 
identify with an automated algorithm. Third, we are interested mainly in 
large-scale shocks produced by mergers and massive accretion of filament gas, 
and visual inspections can easily distinguish them from small-scale shocks
that are abundant. However, such visual identifications may have some potential 
problems. They may miss events of weak or small-scale shocks that are potentially 
interesting. Visual inspections also limit our ability to extend similar 
analysis to a much larger sample. Thus, the results presented in the following sections
should be interpreted with these caveats.

\section{Gas stripping by shocks}\label{sec_re}

Based on the satellites selected { in Section \ref{sec_sample} }, we investigate the stripping and 
compression of the CGM and ISM for ShSATs. 
Subsection \ref{sec_CGM} shows results for the CGM, while the results 
for the ISM is presented in Subsection \ref{sec_ISM}. These results 
for ShSATs are compared with those for NsSATs in Subsection \ref{sec_com}
to demonstrate impacts of shock crossing on the gas contents of satellite 
galaxies.

\subsection{Stripping of the CGM}\label{sec_CGM}

% figure4
\begin{figure*}[htb]
    \centering
    \includegraphics[width=.9\textwidth]{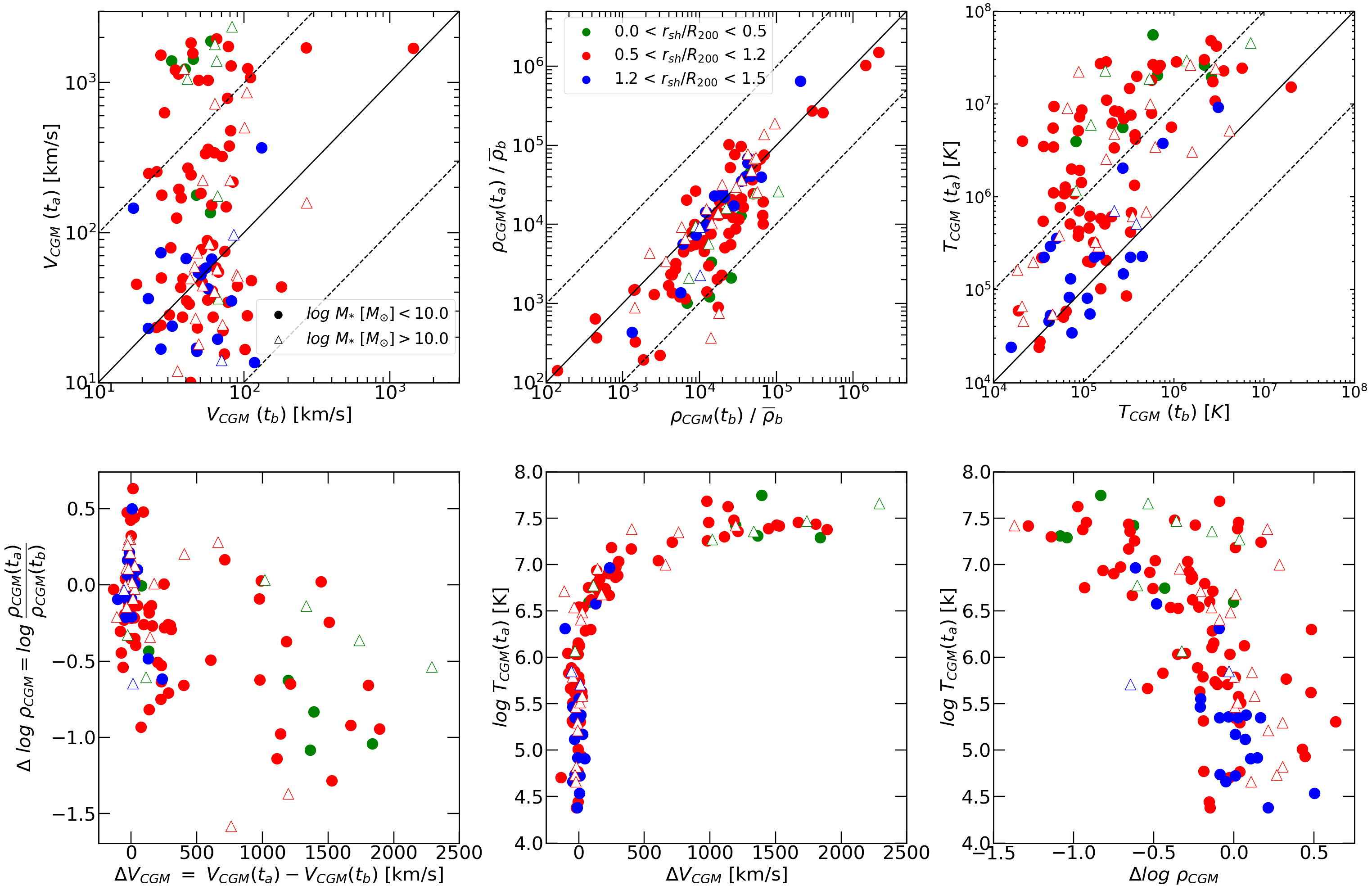}
    \caption{The upper three panels show the comparison of CGM properties, velocity(left), density(middle) and temperature(right) of ShSATs at $t_{\rm b}$ and $t_{\rm a}$. The solid lines show the one-to-one relation and the dashed lines show 1 dex offsets from the solid lines. The lower panels show $\Delta\log\rho_{\rm CGM}$ versus $\Delta v_{\rm CGM}$(left), $T_{\rm CGM}$ versus $\Delta v_{\rm CGM}$(middle)  and  $T_{\rm CGM}$ versus $\Delta\log\rho_{\rm CGM}$. See the text for details.}
    \label{fig_cgmpro} 
\end{figure*}

\begin{figure*}[htb]
    \centering
    \includegraphics[width=.9\textwidth]{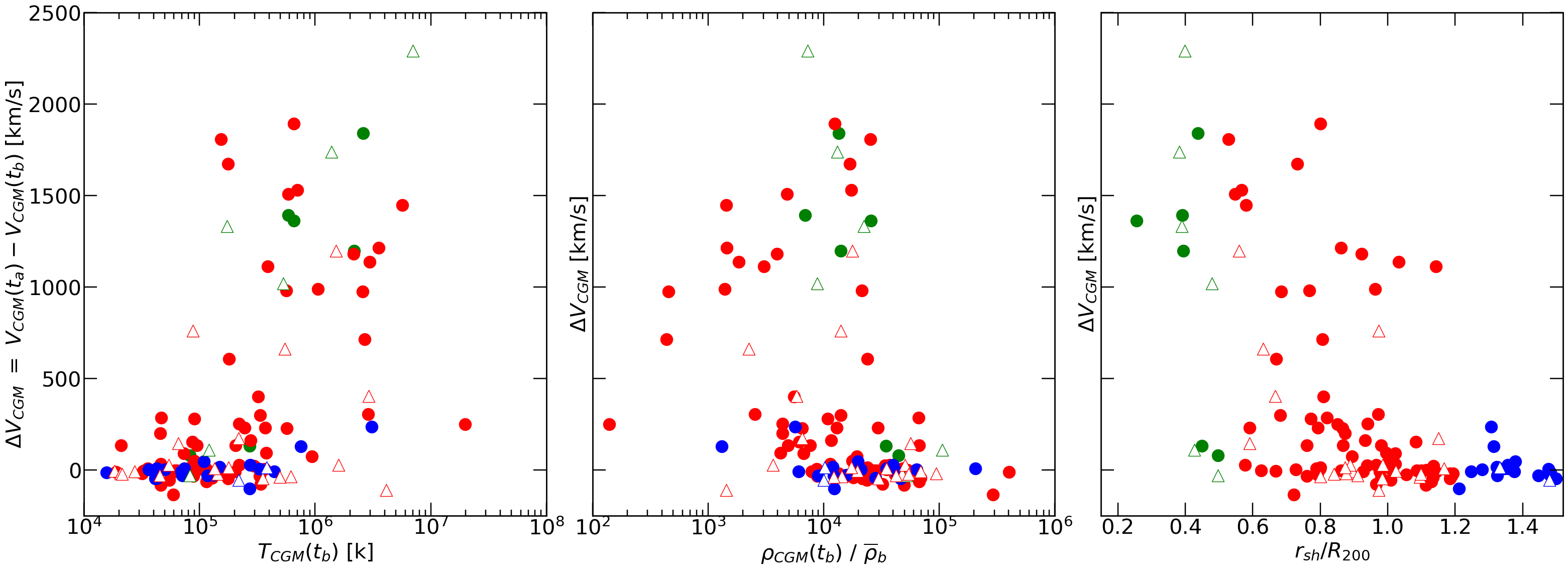}
    \caption{The velocity change, $\Delta v_{\rm CGM}$, as a function of $T_{\rm CGM} (t_{\rm b})$ (left), $\rho_{\rm CGM}(t_{\rm b})$ (middle) and $r_{\rm sh}$ (right). The symbols are the same as in Figure \ref{fig_cgmpro}.}
    \label{fig_dv_reason} 
\end{figure*}

To investigate the impact of shocks on the CGM of ShSATs, 
we first identify the CGM gas cells, defined as the ones which are 
separated from satellites with distances ranging from $2r_{\rm e}$ to 
$4r_{\rm e}$, and then calculate the mean velocity (relative the satellites, $v_{\rm CGM}$), 
mean density ($\rho_{\rm CGM}$) and mean temperature 
($T_{\rm CGM}$) of the CGM gas. Figure \ref{fig_cgmpro} shows the comparison of 
the CGM properties at $t_{\rm b}$ and $t_{\rm a}$, representing times before 
and after the shock-crossing time $t_{\rm sh}$, respectively (see last section).  
Before shock crossing, the CGM velocities are generally low, usually less than 
$100\kms$ relative to the corresponding satellites, much smaller than the typical 
virial velocity of galaxy clusters. This is expected, because the CGM is physically 
associated with the satellite before shock crossing. The CGM density ranges from 
approximately 1000 to $10^5$ times the mean cosmic baryon density 
($\bar\rho_{\rm b}$), consistent with the density in the inner region of a halo 
but much smaller than that of the ISM (typically $>10^6\rho_{\rm b}$). 
The CGM temperature, ranging from $10^{4.5}$ to $10^{6.5}$K, is much lower than 
the virial temperature of a galaxy cluster. These results show that, before
shock crossing, the CGM of most ShSATs has properties expected from the gas 
medium associated with galaxies and their halos. Note that the CGM 
defined here is close to satellites. Thus, our results do not mean that 
CGM at much larger distances is not affected before shock crossing (see the tests for larger scales shown below). 

After interacting with a shock, the CGM velocity changes dramatically. 
Among the 122 ShSATs, 54 of them now have $v_{\rm CGM}(t_{\rm a})>100\kms$, 
28 have $v_{\rm CGM}>500\kms$, and some of them can even reach
a value $\sim 1000\kms$, comparable to the typical virial velocity of a galaxy cluster. 
However, there is a fraction of satellites, for which the CGM velocity is actually 
reduced after shock crossing. Similar behavior 
can also be observed in CGM density and temperature. 
Most satellites have their CGM density significantly reduced after {shock} crossing, 
and the amount of reduction can sometimes reach one dex. 
On average, the mean temperature of the CGM increases by a factor of more than ten.
The lower panels of Figure \ref{fig_cgmpro} show the 
correlations among $\Delta\log\rho_{\rm CGM}$ = $\log(\rho_{\rm CGM}(t_{\rm a})/\rho_{\rm CGM}(t_{\rm b}))$, 
$\Delta v_{\rm CGM}$ = $v_{\rm CGM}(t_{\rm a})-v_{\rm CGM}(t_{\rm b})$ 
and $T_{\rm CGM}(t_{\rm a})$. The strong correlations between these 
quantities indicate that shock crossing simultaneously reduces the CGM density 
and enhances the velocity and temperature of the CGM. 
The correlation between $\Delta v_{\rm CGM}$ and $T_{\rm CGM}(t_{\rm a})$ 
is tighter than other correlations, and it consists of three populations. 
The first one has $\Delta v_{\rm CGM}>500\kms$, with temperature $T_{\rm CGM}(t_{\rm a})
\sim 2.5\times10^7\,{\rm K}$ similar to that of the ICM and almost independent of 
$\Delta v_{\rm CGM}$. As shown in the left panel of Figure \ref{fig_dv_reason},
the median CGM temperature of this population is about $6.7\times10^5\,{\rm K}$
before shock crossing. This temperature is comparable to the virial temperature of 
their subhalos, but much lower than the temperature after shock crossing. 
These satellites have $\Delta\rho_{\rm CGM}$ ranging from 0 to $-1.5$, 
indicating that the original CGM is completely replaced by, or mixed with the 
ICM in their clusters. Thus, the medium around a galaxy at $t_{\rm a}$ does 
not retain its original definition of the CGM that it is bound to its host galaxy.  
For simplicity, however, we still refer such a medium as the CGM.
In the first and second rows of Figure \ref{fig_examples}, we show 
an example (Example one) for this population. As one can see, the CGM 
density drops quickly after interacting with the shock, and 
the temperature changes from $\sim 3.0 \times 10^{6}\,{\rm K}$ at 
$t_{\rm b}$ to $\sim 4.2 \times 10^{7}\,{\rm K}$ at $t_{\rm a}$. 

The second population has $T_{\rm CGM}(t_{\rm a})<10^6$ K. These galaxies 
have $\Delta v_{\rm CGM}$ that are very close to zero, independent of
$T_{\rm CGM}(t_{\rm a})$. The lower-right panel of Figure \ref{fig_cgmpro} 
shows that they have $\Delta\log\rho_{\rm CGM}$ between 0.3 and $-0.5$, with a 
median value close to zero. These results suggest that shock crossing does not 
strip the CGM of these satellite galaxies. However, there are large variations  
of the CGM density between $t_{\rm a}$ and $t_{\rm b}$. For these galaxies, 
although shock crossing cannot strip the CGM quickly, it may compress it. 
Example three shown in Figure \ref{fig_examples} belongs to this population.
We can see a clear bow-like edge surrounding the example galaxy after 
$t_{\rm sh}$, produced by the interaction between the CGM and 
the shock surface. Its CGM is compressed by the interaction so that its density 
increases, but it is still moving with the galaxy. In other cases,
the interaction with the shock surface causes the CGM to disperse and its 
density to decrease, but has not yet stripped the gas.   

The third population has $T_{\rm CGM}(t_{\rm a})>10^6$ K and 
$\Delta v_{\rm CGM}<500\kms$. Different from the first two populations, 
this population exhibits a tight correlation between $T_{\rm CGM}(t_{\rm a})$ 
and $\Delta v_{\rm CGM}$, indicating that the stripping is ongoing. 
In these cases, the gas component between $2r_{\rm e}$ and $ 4r_{\rm e}$ 
at $t_{\rm a}$ is a mixture of the hot ICM and the colder stripped gas.
An example of this population (Example two) is also presented 
in Figure \ref{fig_examples}. Here one can see that the CGM and ISM 
are being stripped and the stripped gas forms an extended tail. 
The stripped ISM in this case enhances the CGM density at $t_{\rm a}$
significantly, indicating that the reduction of CGM density 
may not reflect faithfully the stripping effect. 

A detailed {inspection} of our sample galaxies shows that some satellites are 
severely affected by shocks while others are not. To understand 
the cause of such difference, we show $\Delta v_{\rm CGM}$ versus 
$T_{\rm CGM}(t_{\rm b})$, $\rho_{\rm CGM}(t_{\rm b})$ and $r_{\rm sh}$ 
in Figure \ref{fig_dv_reason}. Here, $\Delta v_{\rm CGM}$ is used to indicate 
the strength of the stripping. As one can see, the first population (with the most 
severe stripping effect) only has $T_{\rm CGM}(t_{\rm b})>10^5$K, 
$\rho_{\rm CGM}(t_{\rm b})<3\times10^{4}\bar\rho_{\rm b}$ 
and $r_{\rm sh}<1.2R_{200}$. In some cases, the CGM gas is actually 
dominated by the extension of spiral arms of galaxies, where
the gas is dense and difficult to strip by the ram pressure.
The dependence on $r_{\rm sh}$ may reflect the fact that the stripping 
efficiency is related to the strength of shocks. It is expected that 
shocks at larger distances are weaker, as their density decreases. 

Other environmental processes, such as tidal stripping, may also lead to 
similar results. To evaluate the contribution of tidal stripping, 
we examined the dark matter mass distribution around satellites, which is
also affected by tidal force. We estimated the dark matter mass in a
spherical shell with radius from $2r_{\rm e}$ to $4r_{\rm e}$. This mass 
is denoted as $M_{\rm dm}$. We then used 
$\Delta\log M_{\rm dm}=\log M_{\rm dm}(t_{\rm a})-\log M_{\rm dm}(t_{\rm b})$ 
to measure the effect of tidal stripping.
We found that $\Delta\log M_{\rm dm}$ is independent of $\Delta v_{\rm CGM}$, 
$\Delta \rho_{\rm CGM}$, and $T_{\rm CGM}(t_{\rm a})$.
Moreover, the median, 16 and 84 percents of the $\Delta\log M_{\rm dm}$ distribution 
are -0.0003, -0.028, and 0.028 dex respectively. These results indicate that tidal 
stripping is not important in the satellite galaxies considered here.

We also investigated the CGM on a larger scale, such as in the range of 
$[4r_{\rm e}, 6r_{\rm e}]$ and $[6r_{\rm e}, 8r_{\rm e}]$. 
The results are quite similar to those shown {in Figure \ref{fig_cgmpro} and Figure \ref{fig_dv_reason}}, so we do not present them here. 
There are three small differences. The first is that, at scales larger than 
$4r_{\rm e}$, a considerable fraction of the CGM is affected by the cluster 
environment even before satellites cross the shock surface.
Second, the CGM on these large scales is more easily affected by shocks 
than that in the inner region, as expected. Third, we found
that the $\Delta M_{\rm dm}$ distribution on these scales is broader than that 
for the inner region, indicating a more important role played by tidal stripping.

\subsection{Stripping of ISM}\label{sec_ISM}

\begin{figure}[htb]
    \centering
    \includegraphics[width=.45\textwidth]{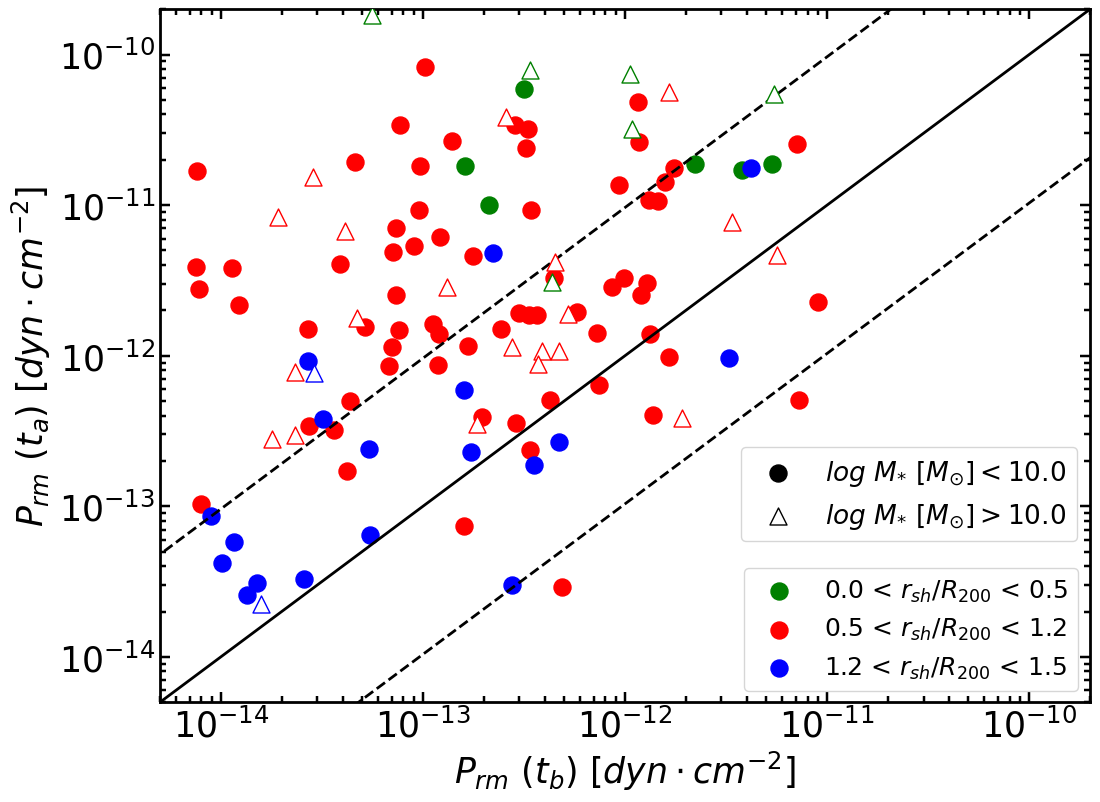}
    \caption{Comparison of the ram pressures on ISM exerted by the surrounding gas at $t_{\rm b}$ and $t_{\rm a}$. The solid line shows the one-to-one relation and the dashed lines show 1 dex offsets from the solid line.}
    \label{fig_rp} 
\end{figure}

\begin{figure*}[htb]
    \centering
    \includegraphics[width=.9\textwidth]{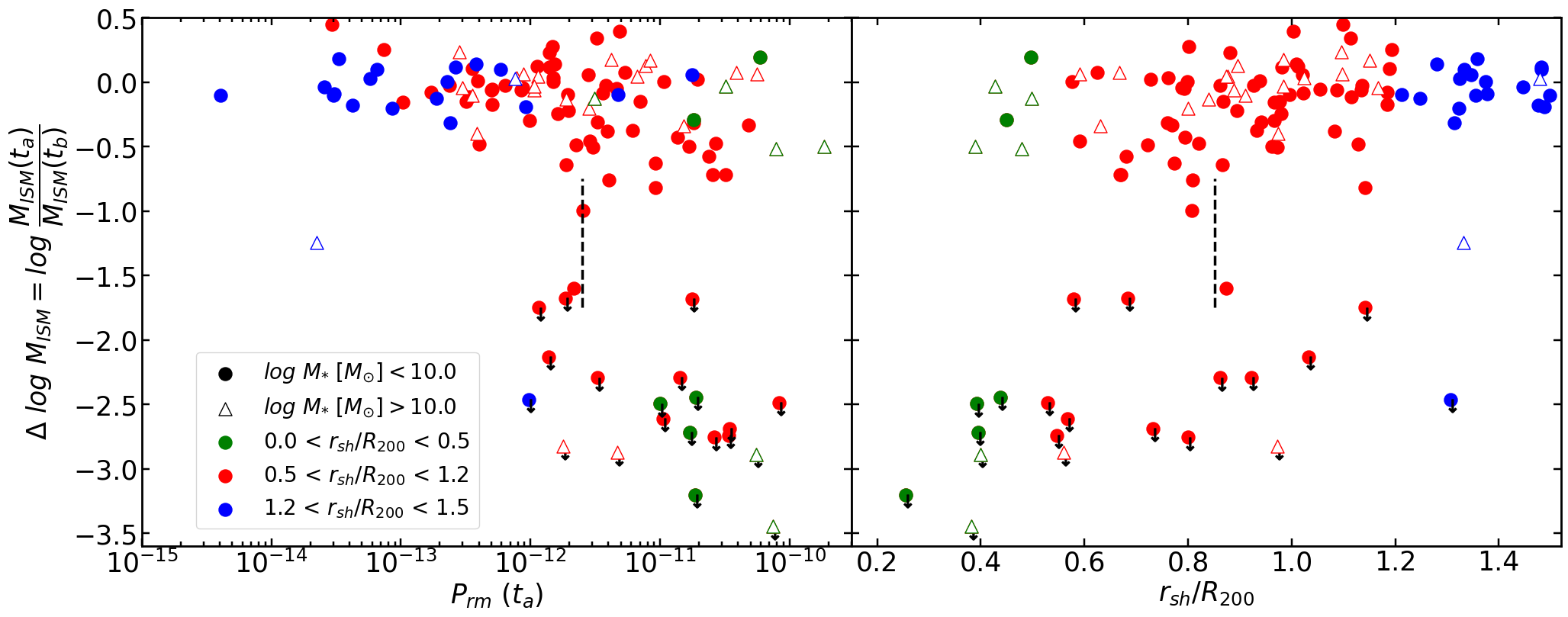}
    \caption{The cold ISM mass loss as a function of the ram pressure at $t_{\rm a}$ (left panel) and $r_{\rm sh}/R_{200}$ (right panel). The symbols with arrows show the upper limit because these galaxies have no cold ISM gas cell at $t_{\rm a}$. One satellite that has cold ISM gas at neither $t_{\rm b}$ nor $t_{\rm a}$ is shown as a dashed line. See the text for details.}
    \label{fig_prmloss} 
\end{figure*}

For our analysis, we define cold gas within $2r_{\rm e}$ in a satellite galaxy
as the ISM gas. Our tests show that before the shock crossing, the cold gas cells 
are closely attached to their host galaxies, with relative velocities 
close to the local rotational velocity. The ram pressure on the ISM is produced 
by its motion relative to the local ICM, and is estimated as
\begin{equation}
    % \mathcolorbox{yellow}{P_{\rm rm}=\rho_{\rm ICM}v_{\rm ICM}^2},\\\label{eq_prm}
    P_{\rm rm}=\rho_{\rm ICM}v_{\rm ICM}^2,\\\label{eq_prm}
\end{equation}
where $\rho_{\rm ICM}$ and $v_{\rm ICM}$ are computed using gas cells in a half spherical shell (with a radius from $6$ to $8r_{\rm e}$) in front of the galaxy moving in the ICM. This radius is chosen 
so that the ram pressure obtained is relevant to the ISM while the influence of any 
extended gas disk (see Figure \ref{fig_examples}) is reduced.  
We present the details of our method and additional tests in Appendix. Figure \ref{fig_rp} compares the 
ram pressure before and after the shock crossing. 
Before shock crossing, the ram pressure is typically less than $10^{-12}\Pre$. 
%As shown in the figure 4 of \cite{Bosseli2022}, $P_{\rm rm}>10^{-12}\Pre$ is required to overcome the gravity of galaxies with masses $9\leq\log(M_*/\Msun)\leq10$, so as to strip their ISM effectively. Thus, most galaxies in our sample can maintain their ISM before shock crossing, i.e. at $t<t_{\rm sh}$. We will come back to this later.
At $t>t_{\rm sh}$, the ram pressure on most of the satellites becomes larger
than that at $t<t_{\rm sh}$, and for some of them the increase is by one to three orders 
of magnitude. A large fraction of satellites have $P_{\rm rm}>10^{-12}\Pre$
at $t>t_{\rm sh}$.
Thus, ram pressure can play an important role in 
stripping the ISM\citep{Bosseli2022}. We cannot see any significant dependence of the ram 
pressure enhancement on stellar mass of galaxies, but a clear dependence on 
$r_{\rm sh}$ can be seen. As shown in Figure \ref{fig_dv_reason}, the CGM is 
affected significantly only at $r_{\rm sh}<1.2 R_{200}$. The change of ram pressure 
is thus expected to be smaller at the larger radius.  After the CGM is stripped, its velocity increases while the density decreases (Figure \ref{fig_cgmpro}). This implies that the boost of the ram pressure is mainly produced by the increase of the velocity of the gas.

\begin{figure*}[htb]
    \centering
    \includegraphics[width=.9\textwidth]{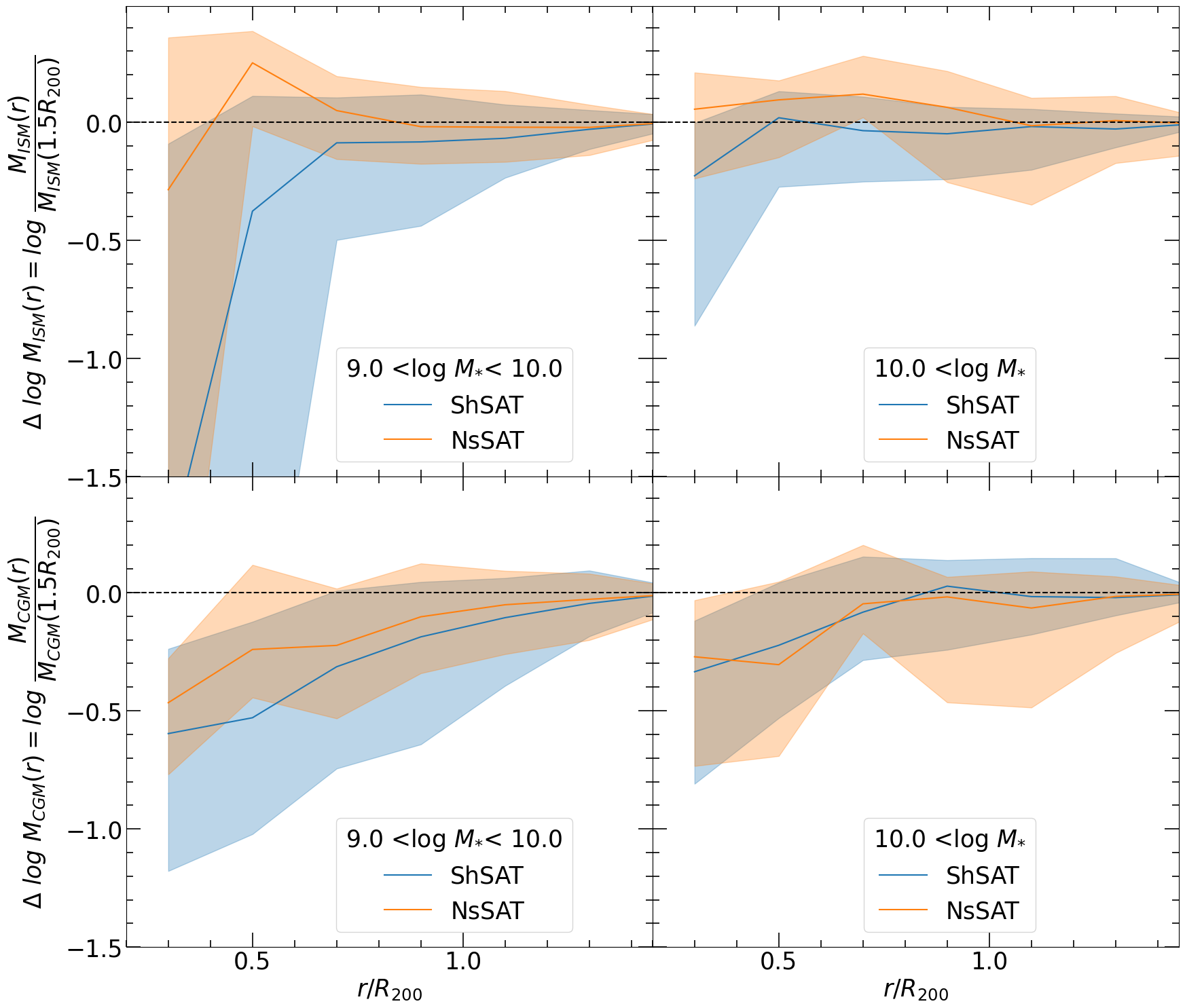}
    \caption{The upper panels show the cold ISM mass change as a function of the halo-centric distance and the lower panels show the results of CGM.
    The left panels are for low-mass galaxies while the right panels are for massive galaxies.
    $\Delta\log M_{\rm ISM}(r)=\log M_{\rm ISM}(r)-\log M_{\rm ISM}(1.5R_{200})$ and $\Delta\log M_{\rm CGM}(r)=\log M_{\rm CGM}(r)-\log M_{\rm CGM}(1.5R_{200})$. The lines and shaded regions show the median, 16 and 84 percents of the distributions. 
    The results for ShSATs and NsSATs are shown in blue and orange respectively.
    For each satellite, we only consider the data before the pericenters of its orbit. Please see the text for the details.}
    \label{fig_gas_r} 
\end{figure*} 

We now investigate the impact of shock crossing on the ISM mass, $M_{\rm ISM}$. Only gas with 
$\log T/{\rm K} <5$ is considered here. Figure \ref{fig_prmloss} shows the change of the ISM mass, 
$\Delta\log M_{\rm ISM}=\log M_{\rm ISM}(t_{\rm a})-\log M_{\rm ISM}(t_{\rm b})$, 
versus the ram pressure at $t_{\rm a}$. For the 20 satellites that do not have cold ISM  
within $2r_{\rm e}$ at $t_{\rm a}$, we set $M_{\rm ISM}$=$1.4\times 10^{6} \Msun$, 
the gas mass resolution of the TNG simulation. We use down-pointing arrows to indicate that the value 
plotted may represent an upper limit. One additional ShSAT has no ISM gas cell at both 
$t_{\rm a}$ and $t_{\rm b}$, and we show it as a dashed vertical line. Our inspection reveals 
that before interacting with the shock, the galaxy has already encountered a 
moving stream and has lost its ISM slowly since then. This is different from the 
other systems which lost their ISM quickly after interactions with a shock. %(see Figure \ref{fig_gas_t}).

In Figure \ref{fig_prmloss}, one can see two distinct populations. 
The first population, consisting of 22 satellites, has a substantial ISM loss 
with $\Delta \log M_{\rm ISM}< -1.0$, and the other population has only small changes 
in the ISM mass, with $\Delta \log M_{\rm ISM}> -1.0$. 
Among the first population,  20 of the 22 satellites have no ISM gas at $t_{\rm a}$, 
indicating complete stripping of the ISM. At $t_{\rm b}$ the median $M_{\rm ISM}/M_{*}$ of 
the 20 satellites is approximately 0.10, and 10 of the 20 galaxies are on the star formation 
main sequence, with specific star formation rates larger than $10^{-10.5}$/yr. 
Most of these galaxies have $P_{\rm rm}> 10^{-12}\Pre$ at $t_{\rm a}$. 
Consistent with the results of \citet{Bosseli2022}, this pressure  
can overcome the gravity of the galaxy and strip the ISM very effectively, 
A clear gap can be seen between the two populations in Figure \ref{fig_prmloss}, 
indicating that the timescale of ram pressure stripping is short so that  
only a small number of galaxies can be observed with parts of their ISM stripped. 
This is consistent with results of some previous studies, which found that
the ISM can be severely stripped over a time scale of 0.1$\sim$0.5 Gyr
as long as the ram pressure overcomes the gravity of the galaxy
\citep[e.g.][]{Quilis2000, Marcolini2003, Tonnesen2009}.
Figure \ref{fig_examples} shows the evolution of the gas density map in an 
example (Example one). Here one can see that the ISM is rapidly stripped after 
it crosses the shock surface. We also show the mass change of the ISM as a 
function of $r_{\rm sh}$ (right panel). As one can see, the first population 
has a very broad distribution in $r_{\rm sh}$, and so the stripping 
of ISM can sometimes be important even at large $r_{\rm sh}$. 
This is different from the predictions of conventional models of ram pressure 
stripping \citep[e.g.][]{Roberts19,Bosseli2022}, in which the effect is expected 
to be important only for small $r_{\rm sh}$. %{\color{red} Is this sentence correct???.}  
For example, as shown in \cite{Bosseli2022}, the predicted stripping-radius for a galaxy with $M_*=10^{9.5}\Msun$ in a 
cluster of $10^{14}\Msun$ is about 1.67$r_{e}$ at $R_{200}$ of the cluster, about $r_{\rm e}$ at 0.5$R_{200}$, and 0$r_{\rm e}$ at 0.2$R_{200}$.
We find two galaxies at $r_{\rm sh}> 1.2R_{200}$, which also have their ISM 
severely stripped. This seems to conflict with the CGM results, which showed that 
the CGM was not significantly affected at $r_{\rm sh}> 1.2R_{200}$. 
Our detailed inspection reveals that one of these two galaxies is massive, and 
its ISM is significantly influenced by internal processes, such as AGN feedback.
The other one is a low-mass galaxy, which is stripped by a strong shock induced by 
a massive merger with a mass ratio of 0.18. The latter case suggests that shocks 
induced by massive mergers can even strip a galaxy located outside the virial 
radius of its host cluster.

\begin{figure*}[htb]
    \centering
    \includegraphics[width=.9\textwidth]{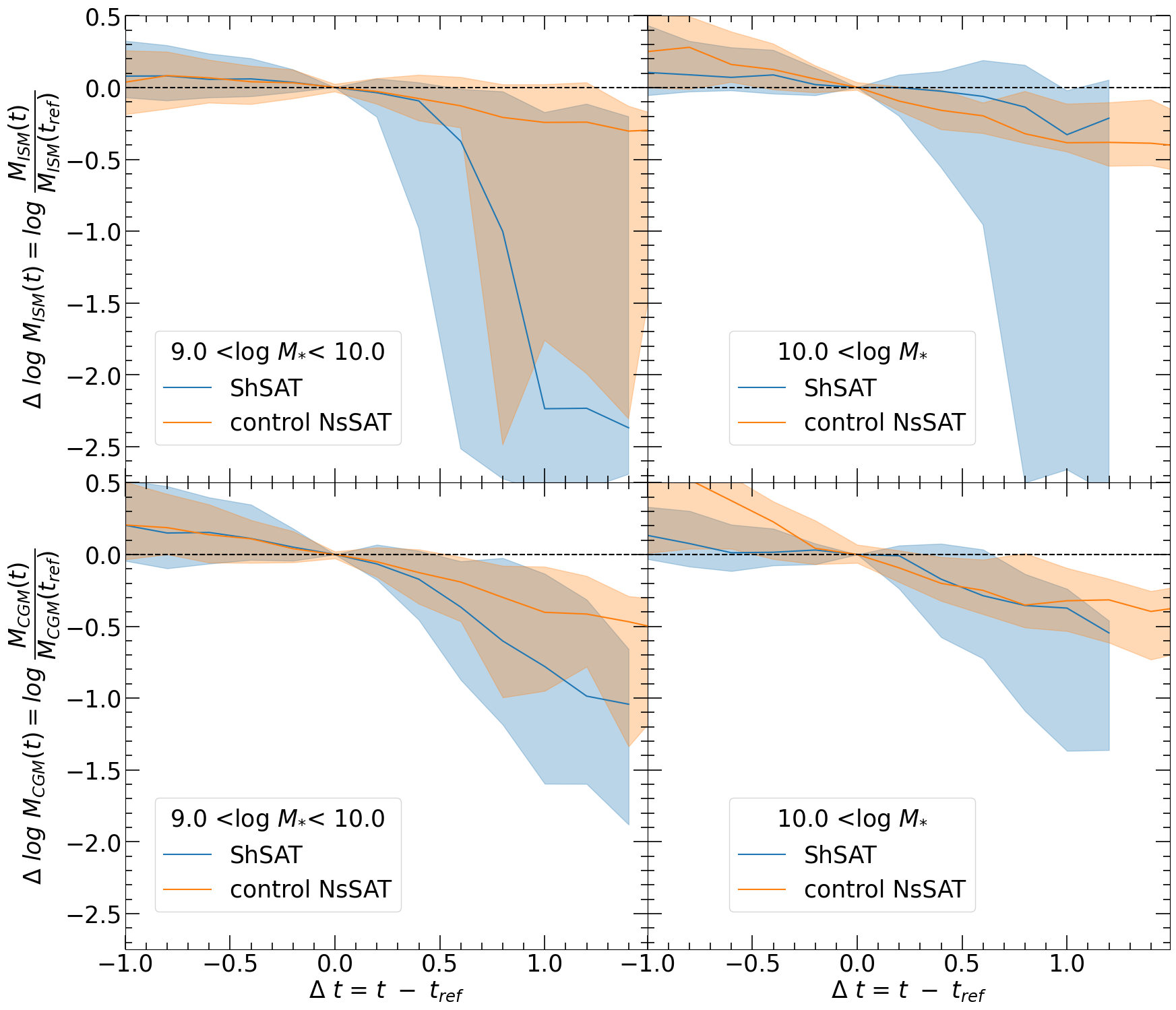}
    \caption{The upper panels show the change in cold ISM mass as a function of time {[Gyr]}, while the lower panels show the results for CGM. 
    The left panels are for low-mass galaxies while the right panels are for massive galaxies.
    $\Delta\log M_{\rm ISM}(t)=\log M_{\rm ISM}(t)-\log M_{\rm ISM}(t_{\rm ref})$ and $\Delta\log M_{\rm CGM}(t)=\log M_{\rm CGM}(t)-\log M_{\rm CGM}(t_{\rm ref})$. 
    Note that $t_{\rm ref}=t_{\rm sh}$ for ShSATs and $t_{\rm ref}=t_{\rm ns}$ 
for control NsSATs.
    The lines and shaded regions show the median, 16 and 84 percents of the distributions. The results for ShSATs and control NsSATs are shown in blue and orange respectively. 
    Note that 5 ShSATs have no $M_{\rm ISM}$ at $t_{\rm sh}$ and are not presented here. For 4 additional ShSATs, we can not find the control galaxies. Please see the text for the details.
    }
    \label{fig_gas_t} 
\end{figure*}

For the second population with small ISM loss, a trend also exists that 
$\Delta \log M_{\rm ISM}$ decreases with increasing ram pressure. We used the 
Spearman rank correlation coefficient to evaluate the strength of the correlation.
The correlation coefficient obtained is $\rho_{\rm r}=-0.35$, with 
$P_{\rm null}= 0.001$. For ShSATs with $r_{\rm sh} < 1.2 R_{200}$, the correlation is stronger, 
with $\rho_{\rm r}=$-0.40 and $P_{\rm null}= 0.001$. The correlation for ShSATs with 
$r_{\rm sh} > 1.2 R_{200}$ is almost zero. Apparently, ram pressure can also affect 
this population of galaxies, particularly those with $r_{\rm sh} < 1.2 R_{200}$, 
although the stripping effect is much weaker than that for the first population. 

There is a wide range of diversity among galaxies. 
Some are fully stripped by shock crossing, as shown by Example one 
in Figure \ref{fig_examples},  while others are affected only slightly, 
as shown by Examples two and three in Figure \ref{fig_examples}.
As discussed in previous studies, effects of RPS depend not only on the density 
and velocity of the surrounding gas considered here, but also on many other factors
\citep[see e.g.][]{Gunn-Gott-72, Hester2006, Bahe2015, Bosseli2022}. 
For example, the orbital shapes, inclinations and intrinsic properties of
satellite galaxies, as well as internal processes, such as feedback, 
can all affect the RPS efficiency, and can produce variances among individual 
galaxies in their suffering of the RPS.

Another important issue is whether or not these galaxies are also 
significantly affected by tidal stripping, instead of RPS. 
To check this, we examined the changes of the stellar mass ($M_*$) and dark matter mass 
$M_{\rm dm}$ within $2r_{\rm e}$. The assumption is that tidal stripping 
also affects stars and dark matter. We found that both $M_{\rm dm}$ and 
$M_*$ remain almost constant from $t_{\rm b}$ to $t_{\rm a}$ for all 
galaxies in our sample. The median, 16 and 84 percentiles of the 
$\Delta\log M_{\rm dm}$ distribution are -0.0023, 0.028, and 0.017 dex, 
respectively, while the values for 
the $\Delta\log M_*$\footnote{$\Delta\log M_*=\log M_*(t_{\rm a})-\log M_*(t_{\rm b})$} 
distribution are 0.0003, -0.0002 and 0.0019, respectively.
These results suggest that tidal stripping may not be important for the loss of 
ISM on the scales interested in here, and are consistent with results 
obtained earlier \citep[e.g.,][]{Bahe2015}.

Figure \ref{fig_gas_r} shows the mass change of the cold ISM as a function of the 
halo-centric distance, $r/R_{200}$. Here, the mass change is 
defined as $\Delta\log M_{\rm ISM}(r)=\log M_{\rm ISM} (r)-\log M_{\rm ISM}(1.5R_{200})$. 
Here the results are presented in terms of the 16\%, 50\%, and 84\% of the 
$\Delta\log M_{\rm ISM}$ distribution. We only consider results before the satellite first reaches 
the pericenter of its orbit. After the pericenter, the distance 
increases with time and the change of ISM mass actually reflects the 
stripping effect at the pericenter, not at the current location.
If the ISM gas of a satellite is completely stripped, we assign 
$M_{\rm ISM}=1.4\times 10^{6} \Msun$, as we did in Figure \ref{fig_prmloss}.
It is thus possible that the results sometimes only reflect upper limits. 
At $r>0.8R_{200}$, low-mass satellites show very small mass change. 
At $r\sim0.8R_{200}$, mass change becomes important for some galaxies, 
and about 25\% of the galaxies have lost more than half of their ISM mass. 
The percentage of galaxies of more than half mass loss increases 
to $\sim 50\%$ at $r=0.5R_{200}$. Almost no distance dependence 
of the mass change is observed for massive galaxies.

We can also investigate the timescale of RPS by shock crossing. 
Figure \ref{fig_gas_t} shows the mass change of the cold ISM, 
$\Delta\log M_{\rm ISM}(t)=\log M_{\rm ISM}(t)/M_{\rm ISM}(t_{\rm sh})$, 
as a function of time ($\Delta t=t-t_{\rm sh}$). Note that 5 ShSATs have lost 
all their ISM at $t_{\rm sh}$ and are excluded from the analysis. 
Before $t_{\rm sh}$, the environmental effect is very weak, 
and there is almost no time dependence of the mass loss for both 
low-mass and massive satellites. This result is consistent with the 
value of the ram pressure before $t_{\rm sh}$ (Figure \ref{fig_rp}),  
which is generally lower than the typical value, $P_{\rm rm}\sim10^{-12}\Pre$, 
required to overcome the gravity of satellites with 
$9\leq\log(M_*/\Msun)\leq10$\citep[see figure 4 in][]{Bosseli2022}.
After $t_{\rm sh}$, low-mass satellites show a quick decline in 
$M_{\rm ISM}$. On average, about half of cold ISM is stripped on a
timescale of 0.6 Gyr after interacting with the shock. About 
one Gyr after $t_{\rm sh}$, sixty percent of the galaxies have lost 
$\sim 90\%$ of their ISM. Since we set the snapshot just before 
shock crossing as $t_{\rm sh}$ and the estimated timescale includes 
that for the stripping of the CGM, the timescale for ISM stripping 
may be significantly shorter than estimated here. 
The stripping effect for massive galaxies is generally weaker 
than that for low-mass galaxies. Only a few high-mass galaxies 
show a significant signature of ISM stripping one Gyr after 
$t_{\rm sh}$, likely because most of the massive galaxies have 
gravitational potential wells deep enough to withstand RPS. 
Detailed inspection shows that these galaxies are also affected 
by shocks, but the impact is weaker, as shown by Example three  
in Figure \ref{fig_examples}. In this particular case, the ram 
pressure compresses the CGM and enhances the ISM density within 
the galaxy, instead of stripping it.

\subsection{Comparison between ShSATs and NsSATs}\label{sec_com}

One important question is whether ShSATs are different from NsSATs in their CGM and ISM. 
Note that NsSATs are also moving in the ICM, although they have not interacted 
with a strong shock front, and so the comparison can provide information 
about the effects of RPS generated by shock crossing.
Figure \ref{fig_gas_r} shows the variation in CGM mass ($\Delta\log M_{\rm CGM}=\log M_{\rm CGM}(r)-\log 
M_{\rm CGM}(1.5R_{200})$) with halo-centric distance, $r/R_{200}$.
As before, the CGM mass is estimated using gas cells within a shell between 
$2-4\ r_{\rm e}$. Similar to the results for ISM,
we only show results for satellites before they reach the pericenters 
of their orbits. We can see a weak but significant trend that the CGM mass 
decreases as satellites move toward the cluster center. This trend exists 
for both low-mass and massive galaxies, although it is weaker for the latter. 
The trend exists for both ShSATs and NsSATs. For low-mass galaxies,
the trend for NsSATs is weaker than that for ShSATs. However, for massive 
galaxies, no significant difference is found between the two populations, 
probably because the sample is too small.

In Figure \ref{fig_gas_r}, we also show the variation in ISM mass as a
function of $r$ for NsSATs in comparison with ShSATs.  For low-mass NsSATs, 
there is almost no radial dependence, and most of the 
ISM can remain as long as $r>0.5R_{200}$. 
When satellites move to the inner region of the cluster ($r<0.5R_{200}$),  
the ISM stripping starts to become important. Again massive galaxies show almost no 
dependence on $r/R_{200}$ in both the inner and outer regions of the clusters.
Note that we do not consider the evolution after the pericenter,  
where a significant ISM loss exists for both ShSATs and NsSATs.
{As discussed in section \ref{sec_ISM}}, ShSATs can be stripped significantly at $r\sim0.8R_{200}$, 
much more effectively than NsSATs, demonstrating that shocks can significantly 
enhance the RPS of the ISM, particularly for low-mass galaxies.

In the previous section, we discussed $\Delta\log M_{\rm ISM}$ as a function 
of $t-t_{\rm sh}$ for ShSATs. For comparison, we also show this dependence 
for NsSATs. Unfortunately, NsSATs do not have a reference time, such as $t_{\rm sh}$. 
We thus design a control NsSAT sample as follows.
For a given ShSAT with $r_{\rm sh}$, we first search for NsSATs with 
a mass similar to the ShSAT (with difference in $M_*$ less than 0.1 dex). 
From these mass-matched galaxies, we randomly select one 
NsSAT whose halo-centric distance has reached $r_{\rm sh}/R_{200}$ 
of the ShSAT at $z\leq0.11$. The selected NsSAT is referred to as the control 
galaxy for the ShSAT. We match both $M_*$ and halo-centric distance to 
eliminate potential dependencies of the gas mass on these two quantities.
The time when the control galaxy first reaches $r_{\rm sh}/R_{200}$ is defined as
the reference time for the control galaxy, and we denote as $t_{\rm ns}$. 
We calculate $t_{\rm ns}$, $M_{\rm CGM}(t_{\rm ns})$, and $M_{\rm ISM}(t_{\rm ns})$ 
by using a linear interpolation between the two adjacent snapshots.
In Figure \ref{fig_gas_t}, we show $\Delta\log M_{\rm CGM}$ and 
$\Delta\log M_{\rm ISM}$ as functions of $\Delta t=t-t_{\rm ref}$, where 
$t_{\rm ref}=t_{\rm sh}$ for ShSATs and $t_{\rm ref}=t_{\rm ns}$ 
for the control sample of galaxies. Here, 
$\Delta\log M_{\rm CGM}(t)=\log M_{\rm CGM}(t)-\log M_{\rm CGM}(t_{\rm ref})$ and 
$\Delta\log M_{\rm ISM}(t)=\log M_{\rm ISM}(t)-\log M_{\rm ISM}(t_{\rm ref})$.
Note that there are 4 ShSATs for which we do not find control galaxies.
Our tests show that excluding the four ShSATs does not change the
results significantly; we include them when showing results for ShSATs.

We first examine the CGM evolution. 
For low-mass ShSATs, there is a weak trend of CGM mass loss at $t<t_{\rm sh}$.
The mass loss is in good agreement with that for the control sample 
at $t<t_{\rm ns}(\Delta t<0)$. At $t>t_{\rm sh}(\Delta t>0)$,
the slope of the $M_{\rm CGM}$ curve for ShSATs becomes steeper, 
suggesting that the interaction with shocks enhances the RPS of CGM. 
However, the slope for the control sample does not change with time. 
The difference leads to a significant deviation between ShSATs and NsSATs 
at $\Delta t>0$. For massive galaxies, however, we see quite different results. 
At $\Delta t<0$, there is a significant difference between the two samples, 
while at $\Delta t>0$, the difference becomes weaker. It is unclear whether 
this reflects the effect of some underlying physical processes or 
it is due to the small sample size (Table \ref{table1}).

A similar but much stronger difference can be found 
in the evolution of the ISM. For both the low-mass and
massive galaxies in the control sample, the median ISM loss usually 
varies slowly with time. Only a small fraction of the low-mass NsSATs 
have prominent ISM loss about one Gyr after the reference time. 
This is broadly consistent with the radial dependence shown in Figure \ref{fig_gas_r}. 
In contrast, ShSATs lose their ISM rapidly after the reference time.
These results suggest that after shock crossing, ShSATs are much more 
affected by RPS than they are before the shock crossing and than NsSATs. 

\section{Summary and discussion}\label{sec_sum}

In this work, we use the IllustrisTNG simulation to study the interaction of large-scale 
shocks with satellites in galaxy clusters at $z\sim0$. The large-scale shocks induced 
by mergers and accretion of gas from massive filaments are common in the local universe. 
Since shock fronts are sharp in the TNG simulation, we can visually identify 
satellite galaxies that have crossed strong shocks. Approximately half of 
star-forming satellites with $\log(M_*/\Msun)\geq 9.0$ have encountered the shocks 
when they move in galaxy clusters. These satellites are referred to as ShSATs.
Satellites that have not interacted with a strong shock are referred to as NsSATs. 
We analyze shock-induced ram pressure stripping (RPS) and other effects on 
satellite galaxies in clusters of galaxies. 

We trace the evolution of the CGM and ISM of the ShSATs and compare 
them with the NsSATs. After shock crossing, the CGM of ShSATs can be 
strongly impacted. In some satellites, the cold and dense CGM is completely 
replaced by the hot and diffuse ICM after shock crossing. 
In others, the CGM is not affected significantly, and in some cases
is even compressed by the shock. After entering the post-shock region,
the ram pressure on the ISM is strongly enhanced in some satellites.
For satellites with $P_{\rm rm}>10^{-12}\Pre$, the ram pressure can overcome
gravity and effectively strip the ISM, particularly for low-mass galaxies 
with $\log(M_{*}/\Msun)<10$.
A significant fraction (50\%) of the ISM can be removed shortly after
the interaction with shocks (less than 0.6 Gyr). Sometimes 
severe stripping of the ISM can occur even in the outskirts of galaxy 
clusters because of the enhancement of ram pressure by shocks. 
The RPS in NsSATs is much weaker than that in ShSATs. Only in the inner 
region of galaxy clusters can the ISM of NsSATs be affected significantly.

Our results suggest that the interaction with large-scale strong shocks 
can enhance RPS. Shocks can accelerate the ICM gas and enhance its density, 
both increasing the ram pressure. Given that a large fraction of satellites 
have interacted with shocks and shocks can affect the properties of the CGM 
and ISM, shocks may play an important role in the evolution of satellite
galaxies, particularly in galaxy clusters, where the ICM is dense, 
merger driven shocks are abundant, and member galaxies are moving at high 
speeds. Our results further suggest that the conventional model, in which 
the ICM is assumed to be static, is inappropriate for modeling the evolution 
of satellite galaxies. 

As we have shown, shock-induced RPS is quite common in present-day clusters of 
galaxies. This may explain that many satellites in the outskirts of galaxy 
clusters show tail-like structure in their cold gas distribution. 
In contrast, the RPS by a static and smooth ICM is expected to be effective
only in the inner regions of clusters. Our results also suggest that a 
large fraction of low-mass satellites can lose their CGM and ISM 
after the emergency of large-scale shocks. Thus, the quenching of star 
formation of satellite galaxies may follow massive mergers that 
drive large-scale shocks. Since large-scale shocks are usually anisotropic, the 
shock-induced RPS may lead to an anisotropic distribution of star-forming and 
quenched populations in clusters. The shock-induced RPS may also impact central 
galaxies if they are swept by shock fronts.  

The sample used in this paper is limited by the volume of the simulation and  
by the visual identification used to identify shock fronts. A large sample with more 
objective selection criteria is needed to carry out a detailed investigation 
of the impact of large-scale shocks on galaxies and its dependence on the 
properties of shocks and the impacted galaxies themselves. We will come back to 
some of these questions in the future.

\section*{Acknowledgements}

We thank the referee for a useful report that significantly improved the paper.
This work is supported by the National Key R\&D Program of China (grant No. 2018YFA0404503), the National Natural Science Foundation of China (Nos.  11733004, 12192224, 11890693, and 11421303), CAS Project for Young Scientists in Basic Research, Grant No. YSBR-062, and the Fundamental Research Funds for the Central Universities. We acknowledge the science research grants from the China Manned Space Project with No. CMS-CSST-2021-A03. The authors gratefully acknowledge the support of Cyrus Chun Ying Tang Foundations. 
The work is also supported by the Supercomputer Center of University of Science and Technology of China.

\bibliography{ref.bib}
\begin{appendix}
 % \section{HOW TO CALCULATE THE RAM-PRESSURE}
In this Appendix, we present details of our method for calculating the ram pressure 
on the ISM of a satellite produced by its surrounding ICM.
According to Equation \ref{eq_prm}, we need both the density ($\rho_{\rm ICM}$) 
and relative velocity ($v_{\rm ICM}$) of the surrounding ICM. For a given satellite, we first 
select all gas cells in a spherical shell with a given thickness. We then calculate the mean 
velocity of the gas shell relative to the satellite. Using the velocity vector so obtained, 
we determine the half spherical shell in front of the satellite. Finally, we use the gas cells 
within this half spherical shell to calculate $\rho_{\rm ICM}$ and $v_{\rm ICM}$. 

\begin{figure*}[htb]
    \centering
    \includegraphics[width=.9\textwidth]{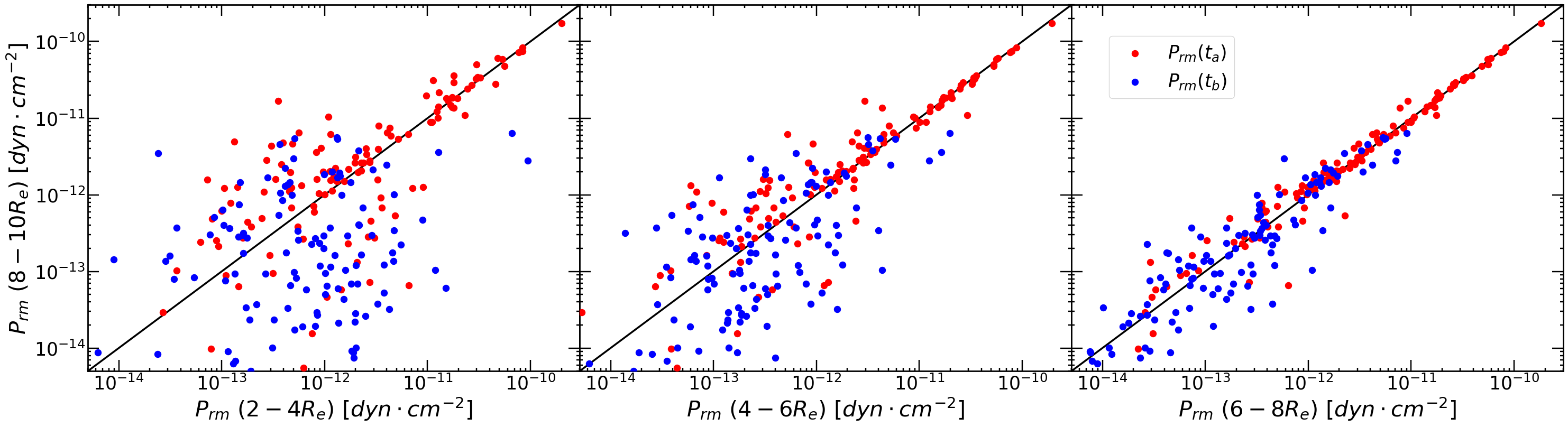}
    \caption{The three panels show the ram pressure calculated using the gas cells at 8-10 $r_{\rm e}$ versus those at 2-4$r_{\rm e}$(left), 4-6$r_{\rm e}$(middle) and 6-8$r_{\rm e}$(right), respectively. The red and blue dots show the results at $t_{\rm a}$ and $t_{\rm b}$, respectively. The solid lines show the one-to-one relation.}
    \label{fig_ap_prm} 
\end{figure*}

One uncertainty in calculating the ram pressure is the radius of the spherical shell
adopted for the calculation. If the radius is chosen to be too small, the 
density and velocity obtained may be affected significantly by the CGM and the ISM. 
On the other hand, if the radius is chosen to be too large,
the estimated ram pressure may not represent well that on 
the ISM of the satellite. In order to find an appropriate scale, 
we perform tests using four different choices for the radius, 
ranging from $r\sim 3r_{\rm e}$ to $r\sim 9r_{\rm e}$. In Figure \ref{fig_ap_prm}
we compare the $P_{\rm rm}$ obtained using $r\sim 9 r_{\rm e}$ (vertical axis) 
with those obtained using smaller $r$.  Note that the ram pressure 
at $t_{\rm a}$ and $t_{\rm b}$ are both presented using different symbols. 

Before analyzing these results, we emphasize that the density and velocity of 
the original CGM bound to a galaxy varies significantly with the distance to 
the galaxy, while the ICM properties are expected to be independent of the 
distance to the galaxy. %as long as the distance is not sufficiently large. {\color{red} I think it should be `sufficiently large', not `not sufficiently large', because you don't want to include CGM which varies rapidly with distance. Please check.} 
Thus, if the gas contents of two different shells are both dominated by the ICM, 
the values of the ram pressure obtained from the two shells 
should be similar. On the other hand, if the gas contents of the two shells  
are dominated by different components or both dominated by CGM, we should expect no correlation or a biased correlation 
between the values of their ram pressure. 

As one can see from Figure \ref{fig_ap_prm}, 
$P_{\rm rm}$(8-10$r_{\rm e}$) and $P_{\rm rm}$(6-8$r_{\rm e}$)
are tightly correlated for both $t_{\rm a}$ and $t_{\rm b}$, and the
values of the ram pressure estimated from the two shells are nearly the same. 
This suggests that at $r> 6 r_{\rm e}$ the gas is dominated by the ICM. 
As $r$ decreases, the difference becomes larger and the correlation 
becomes much looser, indicating that the estimate of the ram pressure 
at smaller $r$ is significantly affected by the CGM. The difference is more significant 
for $t_{\rm b}$ than for $t_{\rm a}$, and at lower $P_{\rm rm}$.
This is expected because the medium at $t_{\rm b}$ or low $P_{\rm rm}$ is
more likely dominated by the original CGM. These tests show that our 
choice of $r=(6-8)r_{\rm e}$ is appropriate for estimating the ram pressure 
from the ICM.
\end{appendix}

% \nolinenumbers %for line number

\clearpage

\end{document}